\documentclass[
    ,final,nomathfonts            
  ]
  {aipproc}

\layoutstyle{8x11single}

\newcommand{\begl}{\begin{equation}}
\newcommand{\engl}{\end{equation}}

\newcommand{\bm}[1]{\mbox{\boldmath$#1$}}
\begin{document}

\title{PCAC and coherent pion production by neutrinos}


\author{Christoph Berger}{
  address={I. Physikalisches Institut, RWTH Aachen university, Germany, email: berger@rwth-aachen.de}
}


\begin{abstract}
Coherent $\pi^+$ and $\pi^\circ$ production in low energy neutrino reactions is discussed in the
framework of the partially conserved axial vector current theory (PCAC). The role of lepton mass effects in suppressing the $\pi^+$ production is discussed. Instead of using models of
pion nucleus scattering,  the available data on pion carbon scattering are implemented for
an analysis of the PCAC prediction. Our results agree well with the published upper limits
for $\pi^+$ production but are much below the recent MiniBooNE result for $\pi^\circ$
production.
\end{abstract}

\maketitle


\section{PCAC and forward lepton theorem}
We discuss single pion production in coherent charged current (CC) and neutral current (NC)
reactions e.g. $\nu_\mu +^{12}\!{\rm C}\rightarrow \nu_\mu +^{12}\!{\rm C}+\pi^\circ$. Our starting point is the general formula for neutrino scattering off a nucleus or nucleon
at rest
\begin{equation}
  \frac{d\sigma^{{\rm CC}}}{dQ^2 dy} = \frac{G_F^2\cos^2\theta_C}{4 \pi^2 } \kappa E
 \frac{Q^2}{|\bm{q}|^2} \left[ u^2 \sigma_L + v^2 \sigma_R + 2uv\sigma_S \right]
  \label{eq:1}
\end{equation}
already derived by Lee and Yang in 1962~\cite{LeeYang} for zero mass of the outgoing
lepton. The momentum and energy transfer between incoming neutrino and outgoing lepton is given by $\bm{q}$ and $\nu=E-E'$.
As usual $Q^2=-q^2$ denotes the four-momentum transfer squared
\footnote{$Q^2=\bm{q}^2-\nu^2;\,y=\nu/E;\, \kappa=(W^2-M^2_N)/2M_N;\, u,v=(E+E'\pm |\bm{q}|)/2E$. $G_F$ and $\theta_C$ are the Fermi coupling constant and the Cabbibo angle.}.
For $Q^2\rightarrow 0$ only the term containing the scalar cross section   $\sigma_S$ survives. 
Here Adler's forward scattering
theorem~\cite{Adler} based 
on PCAC predicts
\begin{equation}
 \sigma_{S,\nu N\to l'F}(W)=\frac{|\bm{q}|}{\kappa Q^2}f_\pi^2\sigma_{\pi N\to F}(W)\enspace .
\label{eq:2}
\end{equation} 
resulting in
\begin{equation}
\frac{d\sigma^{{\rm CC}}}{dQ^2 dy}\bigg|_{Q^2\to 0} = \frac{G_F^2  \cos^2 \theta_C f_\pi^2}{2\pi^2 }\frac{E}{|\bm{q}|}
uv\sigma_{\pi^+ N}(W)
\label{eq:3}\end{equation}
\vspace{-0.5cm } 
and\footnote{$f_{\pi}=\sqrt{2}f_{\pi^\circ}=130.7$ MeV}
\begin{equation}
\frac{d\sigma^{{\rm NC}}}{dQ^2 dy}\bigg|_{Q^2\to 0} = \frac{G_F^2 f_\pi^2}{4\pi^2 }\frac{E}{|\bm{q}|}
uv\sigma_{\pi^\circ N}(W) \enspace .
\label{eq:4}\end{equation}
 For CC the limit $Q^2=0$ cannot be reached. Therefore and for comparison with experiments we extrapolate to 
finite values of $Q^2$ by introducing a formfactor $G_A=m_A^2/(Q^2+m_A^2)$. In addition we include a correction (already contained in Adler's paper) due to the nearby pion pole in the hadronic axial vector current~\cite{Kopeliovich} 
\begl
 \frac{d\sigma^{{\rm CC}}}{dQ^2 dy}= \frac{G_F^2  \cos^2 \theta_C f_\pi^2}{2\pi^2 }\frac{E}{|\bm{q}|}
uv\left[\left(G_A-\frac{1}{2}\frac{Q^2_{\rm min}}{Q^2+m_\pi^2}\right)^2
+\frac{y}{4}(Q^2-Q^2_{\rm min})\frac{Q^2_{\rm min}}{(Q^2+m_\pi^2)^2}\right]\sigma_{\pi^+ N}\enspace .
\label{eq:5}\engl
With $Q^2_{\rm min}=m_{l'}^2y/(1-y)$ the pion pole term vanishes for $m_{l'}=0$, it is a lepton mass correction.
\section{Coherent scattering}
\begin{figure}
\includegraphics[height=0.3\textheight]{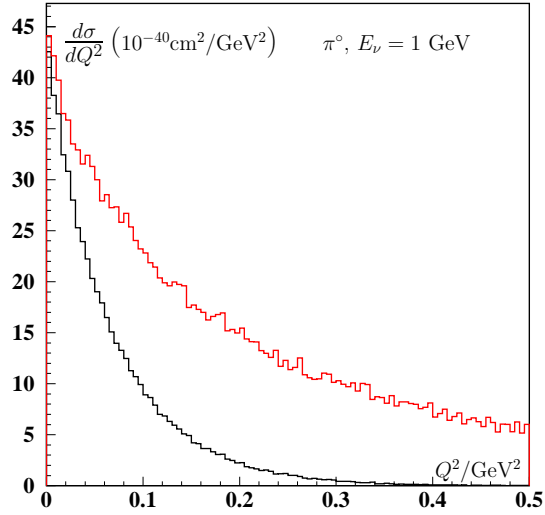}
\caption{$\pi^\circ$ production by neutrino scattering on carbon. Black histogram calculated from integrating  (\ref{eq:6a}) over $(t,y)$, red histogram from
integrating  (\ref{eq:4}) over $y$. A hadronic toy model is used
(see text).}
\label{fig1}\end{figure}

Coherent pion nucleus ($\pi N$)  scattering is strongly peaked in forward direction
distinguishing it from incoherent background. We therefore expect coherent single pion production by neutrinos to be well described by the PCAC ansatz. Like in the original Rein Sehgal (RS) paper~\cite{RS1} this approximation is assumed to hold also for the 
differential cross section 
\begl
 \frac{d\sigma^{{\rm CC}}}{dQ^2 dy dt}=\frac{G_F^2  \cos^2 \theta_C f_\pi^2}{2\pi^2 }\frac{E}{|\bm{q}|}
uv\left[\left(G_A-\frac{1}{2}\frac{Q^2_{\rm min}}{Q^2+m_\pi^2}\right)^2
+\frac{y}{4}(Q^2-Q^2_{\rm min})\frac{Q^2_{\rm min}}{(Q^2+m_\pi^2)^2}\right]
\frac{d\sigma (\pi^+ N\to\pi^+ N)}{dt}
\label{eq:6}
\engl
and
\begl
 \frac{d\sigma^{{\rm NC}}}{dQ^2 dy dt}=\frac{G_F^2 f_\pi^2}{4\pi^2 }\frac{E}{|\bm{q}|}
uv
\frac{d\sigma (\pi^\circ N\to\pi^\circ N)}{dt}\enspace .
\label{eq:6a}
\engl
This extension is by no means trivial. $t$ is the four momentum transfer squared
between the incoming virtual boson and the outgoing pion. Therefore $t=0$ cannot be reached.
$t_{\rm min}=f(Q^2)$ results in a very effective $Q^2$ cutoff for exponentially decreasing hadronic differential cross sections.
The $t$-integral of e.g. 
(\ref{eq:6}) approaches (\ref{eq:5}) only for $Q^2\to 0$. Figure(\ref{fig1}) shows as example $\pi^\circ$ production on carbon
for $E=1$ GeV.  A hadronic toy model $d\sigma/dt=a\exp(-bt)$ with constant coefficients  $a=3200\, {\rm mb/GeV}^2$,
$b=40\, {\rm GeV}^{-2}$ is used.
\begin{figure}[h]
\includegraphics[width=0.46\textwidth]{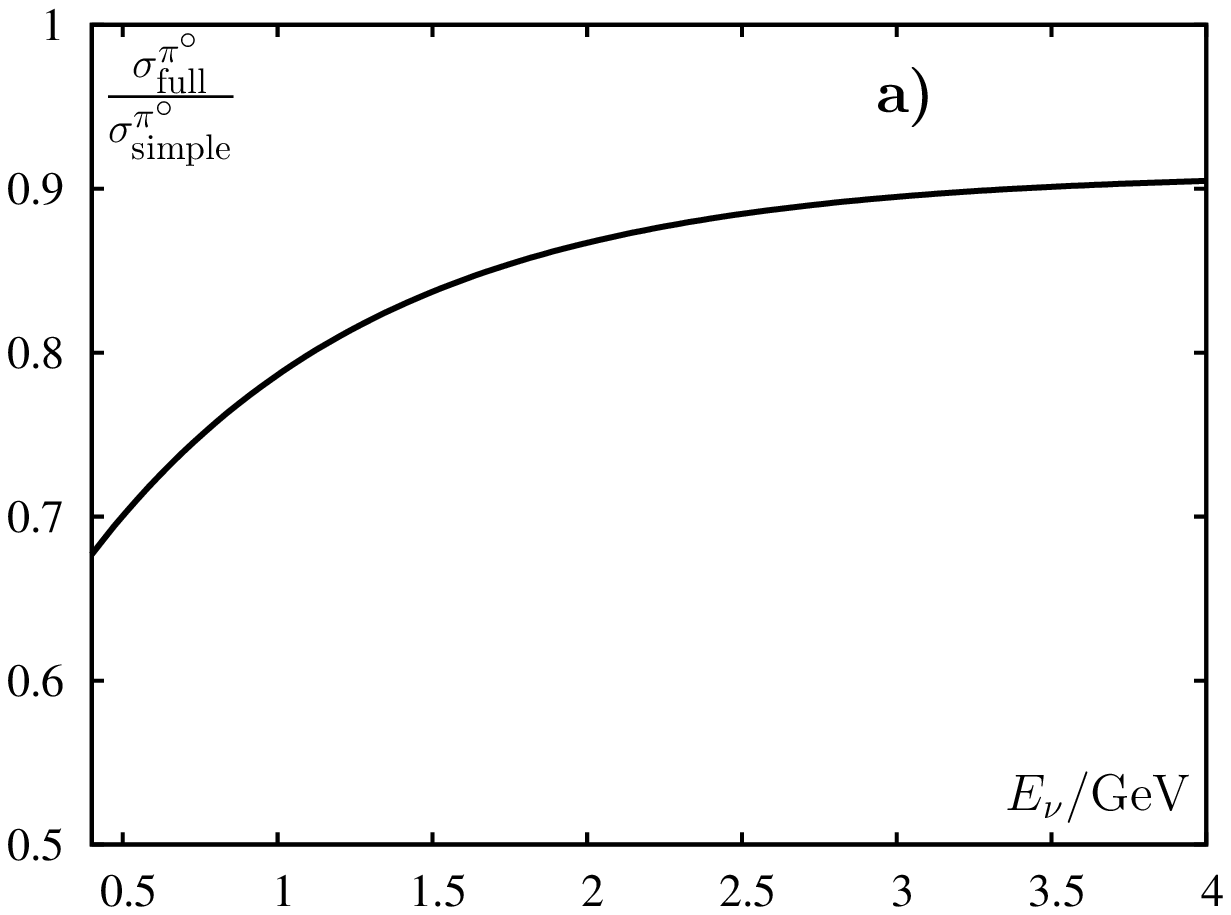}
\includegraphics[width=0.46\textwidth]{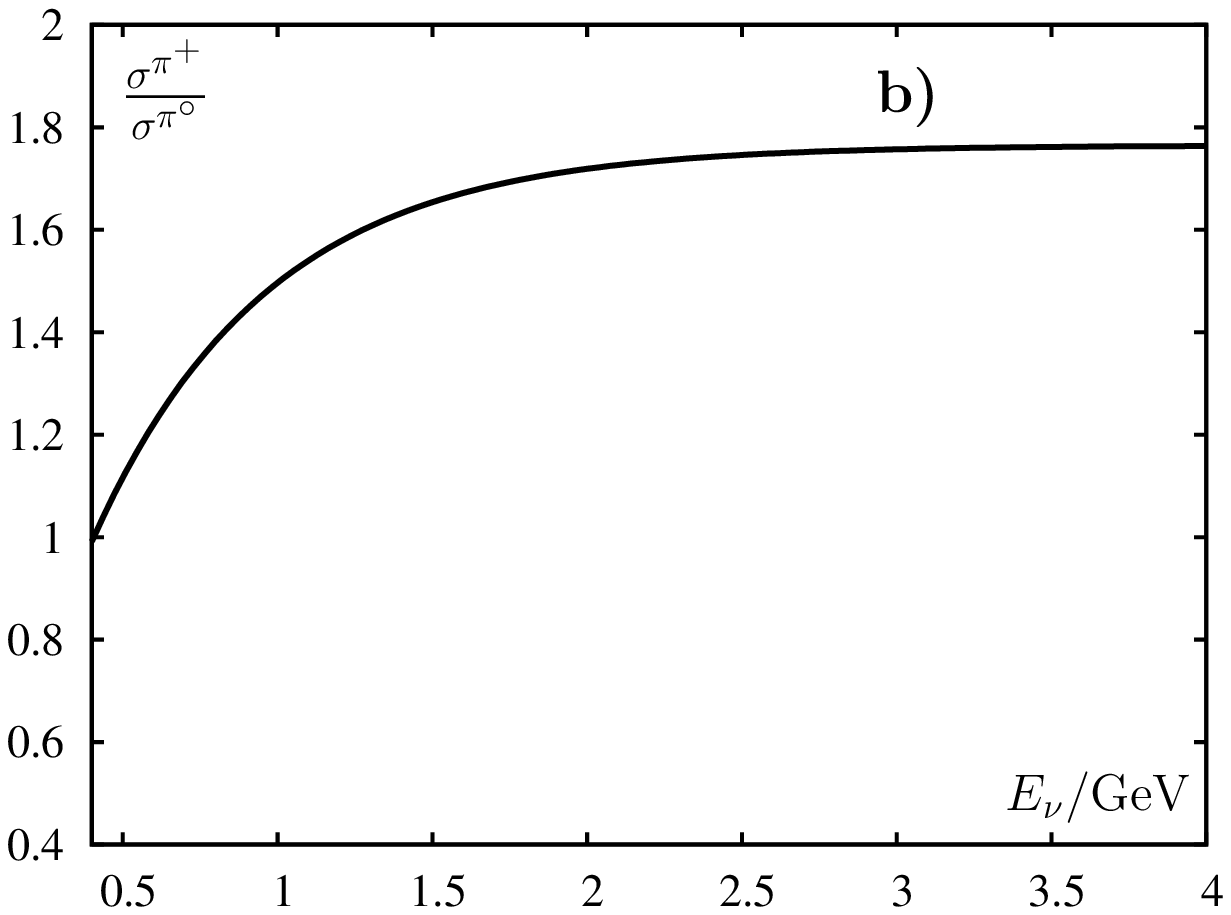}
\caption{a) Ratio $\sigma^{\pi^0}_{\rm full}/\sigma^{\pi^0}_{\rm simple}$ 
of the integrated cross sections of (\ref{eq:6a})  versus  
the energy of the incoming neutrino. $\sigma^{\pi^0}_{\rm simple}$ is calculated using the kinematical approximations of~\cite{RS1}. b)  Ratio $\sigma^{\pi^+}/\sigma^{\pi^0}$ 
of the integrated cross sections  versus 
the energy of the incoming neutrino. }
\label{fig2}
\end{figure}  

The RS paper~\cite{RS1} evaluates the kinematical factor always at $Q^2=0$, i.e. 
$Euv/|\bm{q}|\to (1-y)/y$. At high energies the differences are negligible. At threshold
they are very important. This is demonstrated in figure (\ref{fig2}a) using
the hadronic toy model. The CC/NC ratio shown in figure (\ref{fig2}b) approaches the
limiting value of $2\cos^2 \theta_C$ also only at high energies.

\section{The elastic pion nucleus cross section}
A simple model for elastic pion nucleus scattering, which can be 
easily implemented into MC generators is also contained in~\cite{RS1}. We discuss it here for
isoscalar targets of atomic mass $A$ leading apart from electromagnetic corrections to identical
cross sections for $\pi^{\pm,\circ}$. Starting from
\begin{equation}
 \frac{d\sigma(\pi N\to\pi N) }{dt}=A^2\frac{d\sigma_{\rm el}}{dt}\Big|_{t=0}e^{-b_{\rm RS}t}F_{\rm abs}\enspace 
\label{eq:10}
\end{equation}
the elastic differential pion nucleon cross section at $t=0$ is calculated with the help of the optical theorem
\begin{equation}
 \frac{d\sigma_{\rm el}}{dt}\Big|_{t=0}=\frac{1}{16\pi}
\left(\frac{\sigma^{\pi^+p}_{\rm tot}+\sigma^{\pi^-p}_{\rm tot}}{2}\right)^2
\end{equation}
where the total pion proton cross sections are taken from data.
The slope $b_{\rm RS}$ is determined via the optical model relation
\begin{equation}
 b_{\rm RS}=\frac{1}{3}R_0^2A^{2/3}\,\,\, {\rm e.g.}\,\, R_0=1.057\,{\rm fm}\enspace .
\end{equation}
Finally using a simple geometrical picture the absorption factor
\begin{equation}
 F_{\rm abs}=\exp{\left(-\frac{9A^{1/3}}{16\pi R_0^2}\sigma_{\rm inel}\right)}
\label{eq:13}\end{equation}
is calculated from data for inelastic pion proton scattering via
\begin{equation}
\sigma_{\rm inel}=\frac{\sigma^{\pi^+p}_{\rm inel}+\sigma^{\pi^-p}_{\rm inel}}{2}\enspace .
\end{equation}
Although this model has its limitations (e.g. it predicts
$d\sigma/dt\to 0$ for $A\to \infty$) it has been very successful in describing high energy
coherent neutrino scattering~\cite{Kopeliovich}.

\begin{figure}
\includegraphics[width=0.46\textwidth]{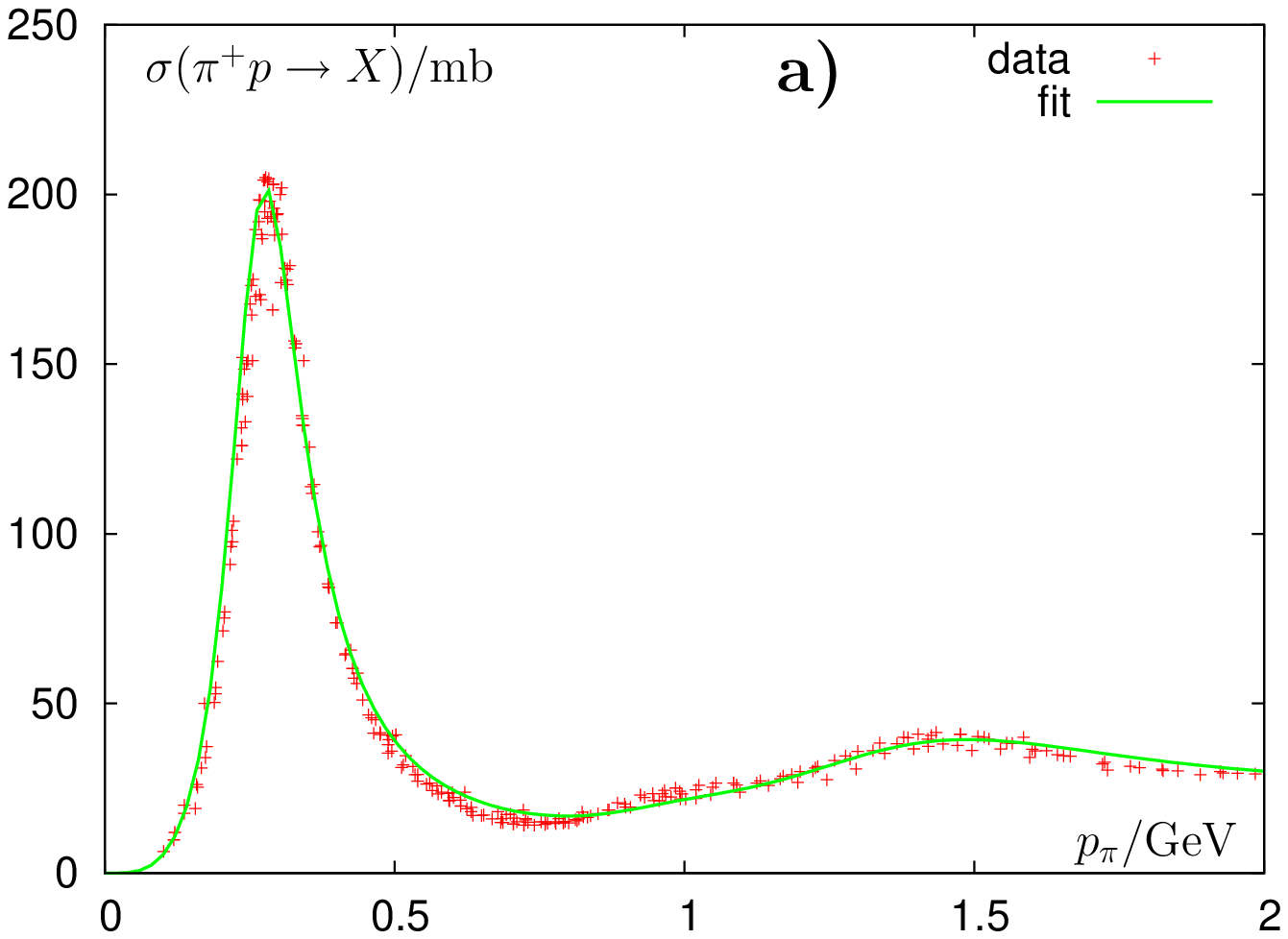}\hspace{0.2cm}
\includegraphics[width=0.46\textwidth]{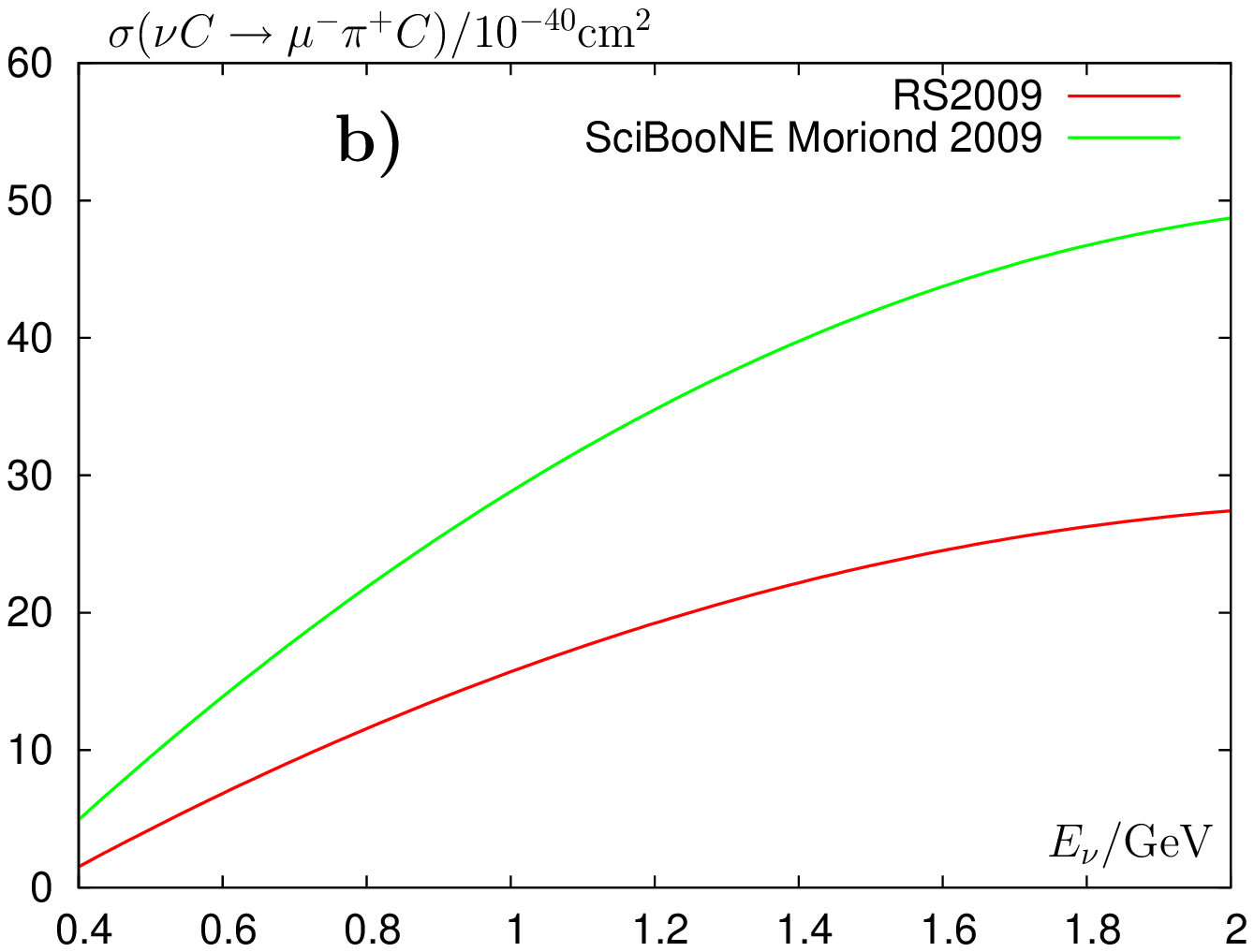}
\caption{a) Fit to the total $\pi^+p$ cross section using the PDG tables~\cite{pdg} for pion laboratory momenta up to
2 GeV. b) Total CC cross section versus neutrino energy using the updated hadronic
RS model (RS2009). For comparison the prediction used by the SciBooNE Collaboration~\cite{Sci}
is also shown.}
\label{fig3}\end{figure}

In order to obtain a more precise prediction for the pion nucleus cross section in the resonance
region  the
parameterization of the pion nucleon cross sections used in RS~\cite{RS1} has been replaced
by detailed fits to the $\pi^\pm p$ data published by the Particle Data Group~\cite{pdg}.
An example is shown in figure (\ref{fig3}a). The curve labelled RS2009 in figure (\ref{fig3}b)
shows the resulting
total CC coherent neutrino cross section for energies up to 2 GeV.
There exist various implementations of the RS-model
which, however, obtain different results. An example is displayed in figure (\ref{fig3}b). The
predictions of other Monte Carlo generators claiming to use the RS-model show even more pronounced discrepancies~\cite{Dytman}. The reason for these differences remains a puzzle.

\begin{figure}
\includegraphics[width=0.46\textwidth]{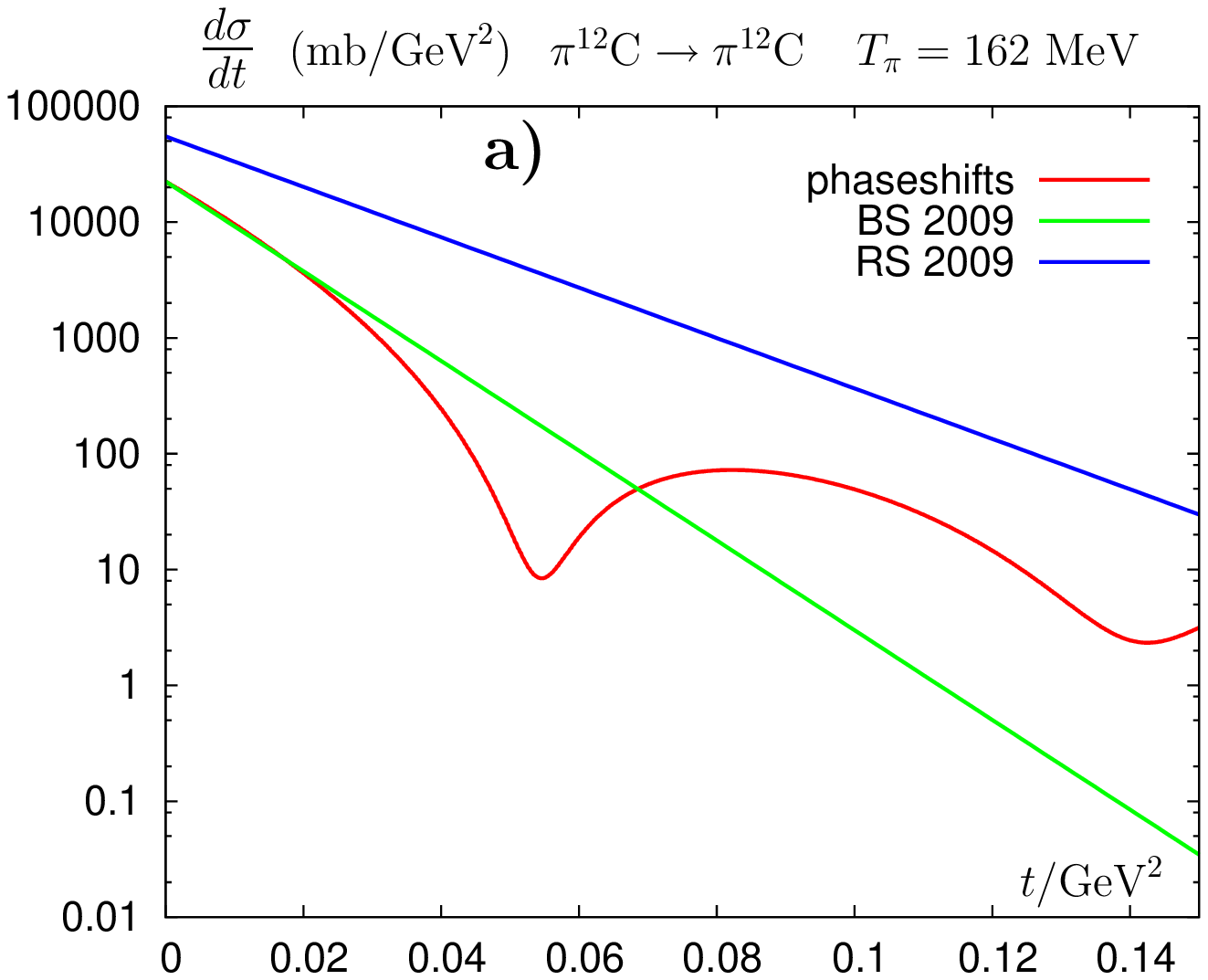}\hspace{0.2cm}
\includegraphics[width=0.48\textwidth]{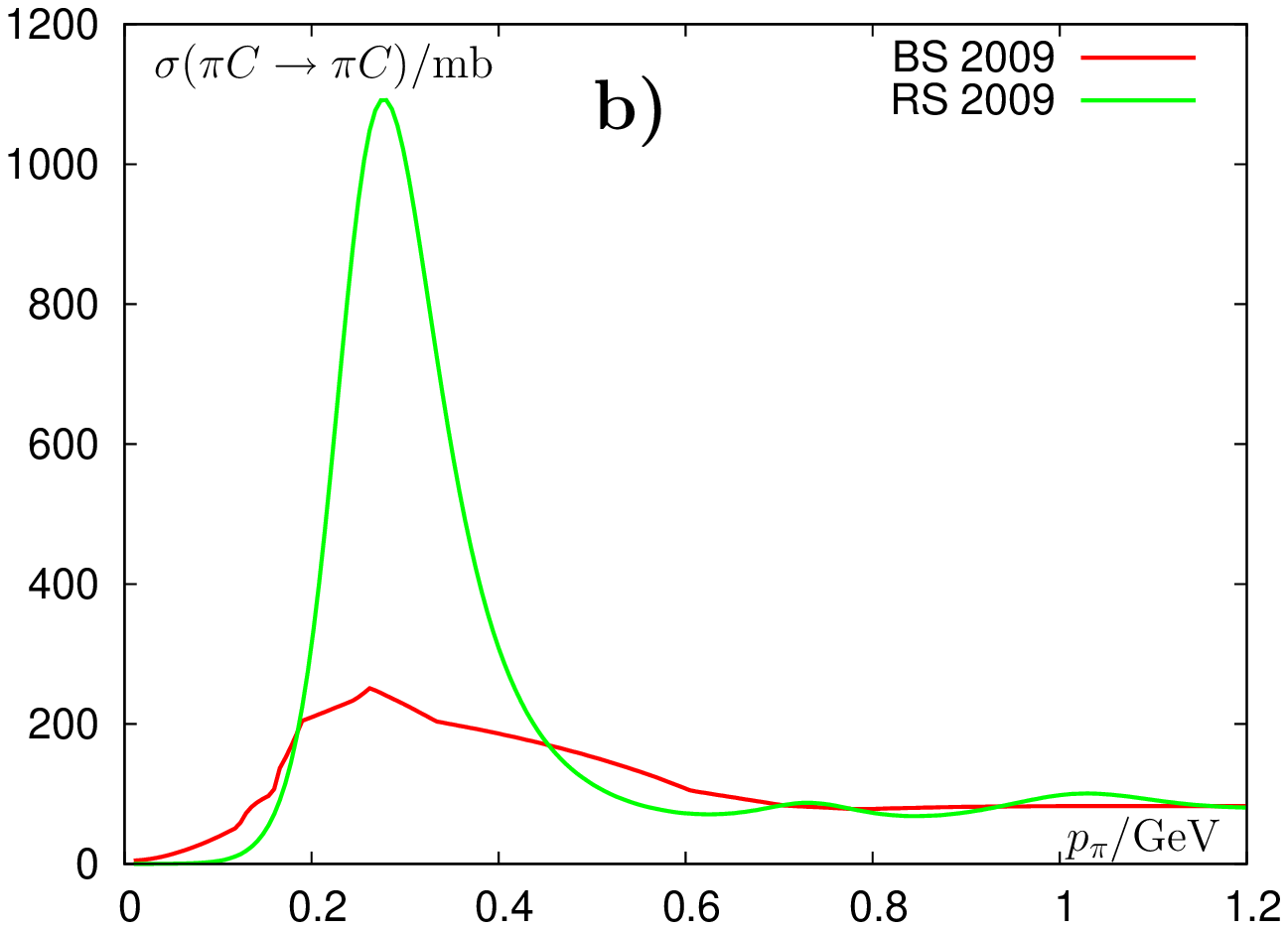}
\caption{a) Differential cross section $d\sigma/dt$ for elastic pion carbon scattering.
Th red line labelled 'phasehifts' is calculated from the phasehifts in~\cite{Karlsruhe}.
The green line (BS2009) is an exponential fit to the forward cross section. The blue
line (RS2009) represents the result of the updated RS hadronic model. b) Total elastic pion carbon cross
section versus the laboratory pion momentum in the updated RS hadronic model (RS2009) and from
exponential fits to the phaseshift analysis (BS2009).}
\label{fig4}\end{figure}

For the resonance region with its rapidly varying cross sections and angular distributions
the hadronic RS-model is probably too simple. It describes badly the low energy 
experimental data on elastic
$\pi ^{12}\!{\rm C}$ scattering. Instead of refining it we -- in the spirit of Adler's theorem -- directly revert to
the measured $\pi ^{12}\!{\rm C}$  cross sections. 
Pion carbon scattering data with $30 <T_\pi<776\,{\rm MeV}$ of various experiments  have been subjected to a phase shift analysis
and extrapolated to $T_\pi=870$ MeV by the Karlsruhe group~\cite{Karlsruhe}. Figure (\ref{fig4}a) shows an example for the differential cross section $d\sigma/dt$ reconstructed from these phaseshifts at
$T_\pi=162$ MeV, close to the maximum of the first resonance. 
The forward scattering containing the bulk of the cross section is then fitted by a $a\exp{(-bt)}$ ansatz resulting in energy dependent coefficients
$a,b$~\cite{BS}.
Using this parameterization the pion carbon elastic cross section in the resonance region 
is below the hadronic RS-model but approaches it quickly at higher $p_\pi$, see figure (\ref{fig4}b).

\section{Results}
\begin{figure}[h]
\includegraphics[width=0.46\textwidth]{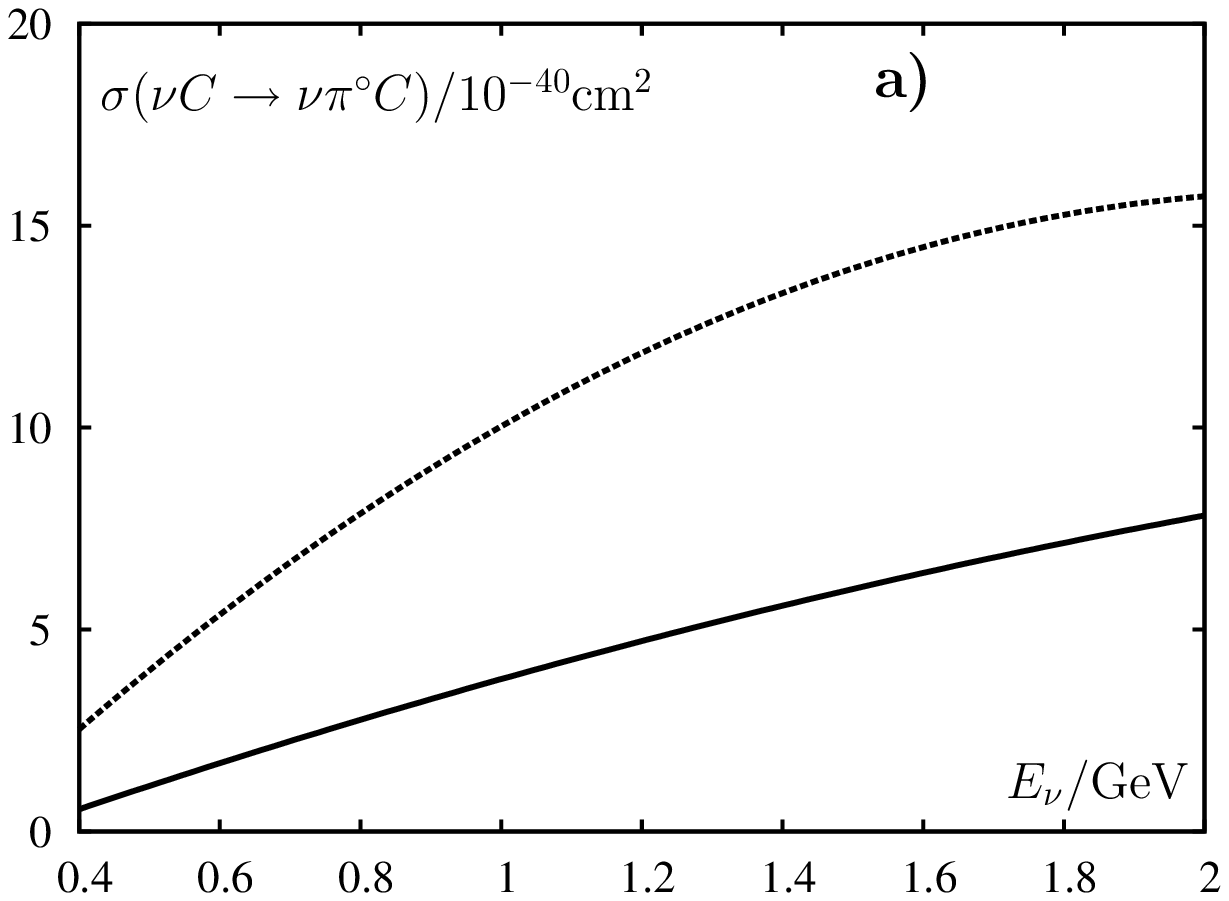}\hspace{0.2cm}
\includegraphics[width=0.46\textwidth]{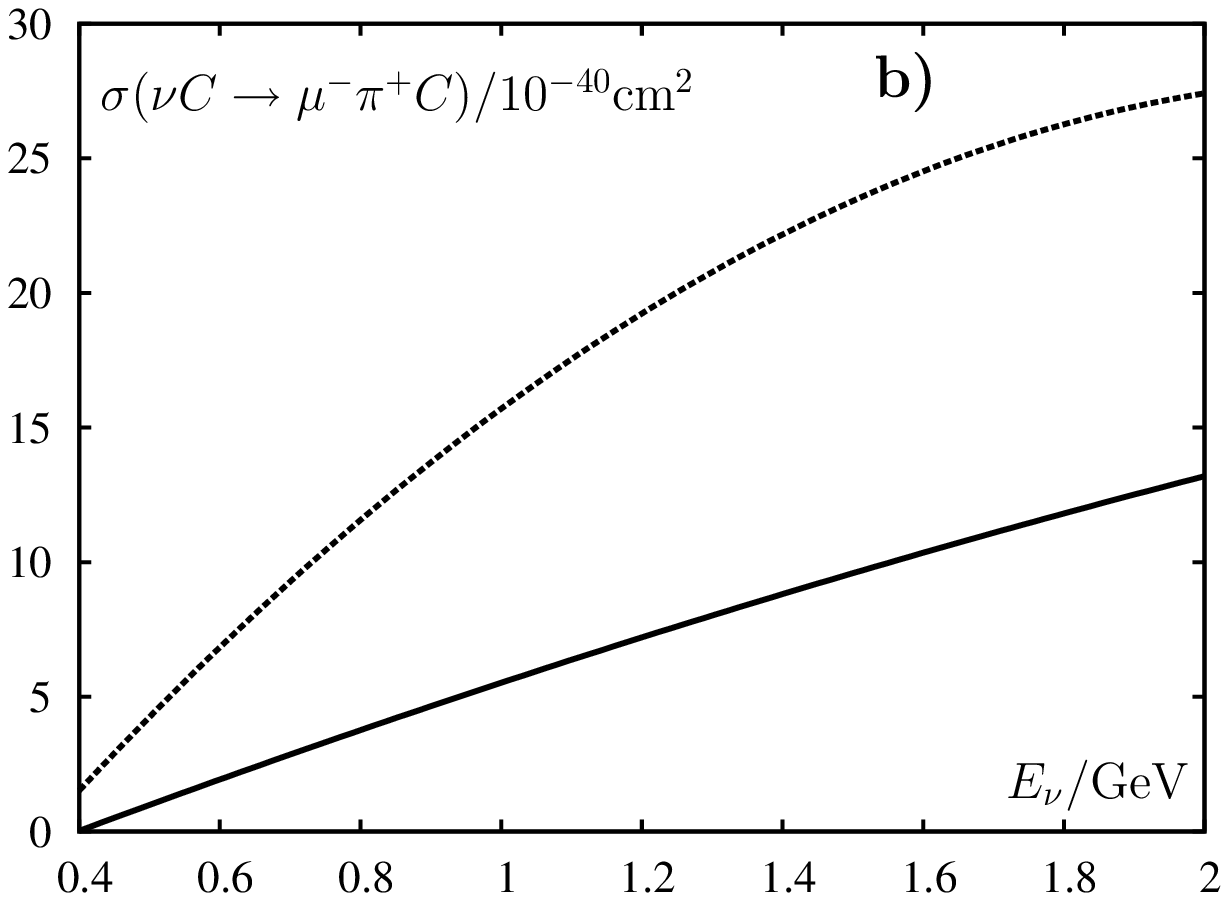}
\caption{Cross section per nucleus of coherent $\pi$ production
by neutrinos off carbon nuclei, a) NC reaction $\nu_\mu +^{12}\!{\rm C}\rightarrow \nu_\mu +^{12}\!{\rm C}+\pi^\circ$, b) CC reaction $\nu_\mu +^{12}\!{\rm C}\rightarrow \mu^- +^{12}\!{\rm C}+\pi^+$.
The data in units of $10^{-40}$ cm$^2$
are plotted versus the neutrino energy in GeV. The upper curve is calculated using
the hadronic RS model, the lower curve using our parametrization
of pion carbon scattering data.}
\label{fig5}\end{figure}

Using the fits discussed in the preceding section we  get a substantial modification of the PCAC prediction for pion production
off carbon nuclei for NC and CC reactions at low neutrino energies. This is demonstrated in figure (\ref{fig5})
where the new results are compared with calculations using the updated hadronic RS-model.

\begin{figure}[h]
\includegraphics[width=0.43\textwidth]{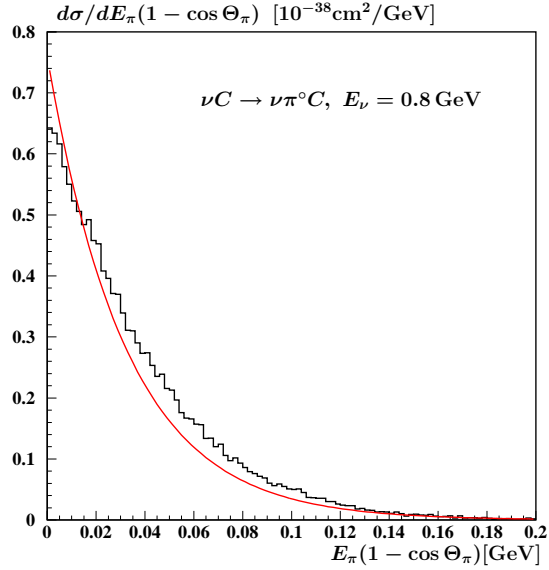}
\caption{Differential cross section $d\sigma/dE_\pi(1-\cos\Theta_\pi)$ versus $E_\pi(1-\cos\Theta_\pi)$
calculated fur NC pion production at a neutrino energy of 0.8 GeV. The histogram uses the
BS2009-model~\cite{BS} and the curve represents the results of a recent nuclear physics 
calculation~\cite{Nieves}. }
\label{fig6}\end{figure}
Our predictions for the total cross section  are compatible with other PCAC based
calculations~\cite{Paschos}
and remarkably close to certain variants of
microscopic nuclear physics models~\cite{Singh, Amaro}.
Differential distributions are more sensitive to model details. The MiniBooNE collaboration has
proposed to use $E_\pi(1-\cos\Theta_\pi)$ as variable for the analysis of neutrino scattering~\cite{MbooNE}
($E_\pi$ and $\Theta_\pi$ defined in the laboratory system).  As can be seen in in figure~(\ref{fig6})
a recent  nuclear physics model~\cite{Nieves,Amaro} agrees well with our PCAC model at a neutrino energy
typical for the MiniBooNE experiment. 

The extension of the new ansatz to other nuclei is of particular importance.
At this moment we propose to use in the spirit of the optical model  an $A^{2/3}$ scaling law which is close
to the effective $A$-dependence obtained in the hadronic RS-model for light nuclei.

Our results agree with the published experimental limits on coherent $\pi^+$ production~\cite{K2K,Sci1}.
Using the $A^{2/3}$ scaling law it also agrees with the $\pi^\circ$ data of the Aachen Padova experiment~\cite{AP}.
Like all other recent theoretical predictions we have a problem with the MiniBooNE $\pi^\circ$ result~\cite{MbooNE}. PCAC models
have very little flexibility in tuning the predictions.  Before, however, claiming that
the  model is falsified one would like to clarify several questions, e.g.
the dependence of the experimental result on the use of inappropriate Monte Carlo models and 
how the experiments ensure the coherence of the process.
\begin{figure}[h]
\includegraphics[width=0.43\textwidth]{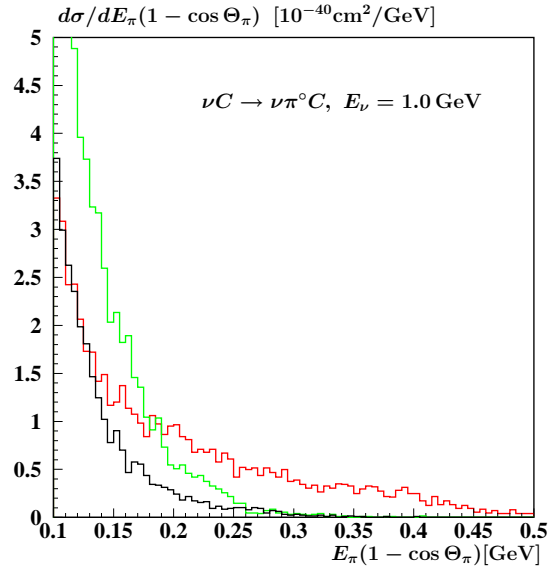}
\caption{Tail of the differential cross section $d\sigma/dE_\pi(1-\cos\Theta_\pi)$ versus $E_\pi(1-\cos\Theta_\pi)$
calculated for NC pion production at a neutrino energy of 1.0 GeV. The black histogram is
obtained using  the two parameter fit to pion carbon scattering described in~\cite{BS}.
The red histogram uses the full angular distribution. For comparison a prediction
using the updated hadronic RS-model is also shown (green histogram).}
\label{fig7}\end{figure}

Instead of using a two parameter fit of the pion carbon differential cross section a variant
of the new PCAC model has been studied in which the full angular distribution as represented
by the phase shifts (see figure (\ref{fig4}a)) is utilized. For energies between two data sets a
linear extrapolation of the differential cross sections $d\sigma/dt$ at a fixed scattering
angle in the CMS system is applied. The resulting total neutrino cross section and the
pion angular distribution in the forward direction changes only at the  level of
 a few percent. In contrast to the simpler model there is, however, a long tail at larger
angles. An example is shown in figure (\ref{fig7}). These differences might become important
when precise experimental data are available. With only a few sets of phaseshifts for
other nuclei on-hand the  extension of this model version to non carbonic targets requires further research.





\begin{theacknowledgments}
This paper is a writeup of a talk presented at NUINT 2009, the sixth international workshop on neutrino interactions
in the few GeV region in May 2009 at Sitges, Spain. The excellent work of the organizers of NUINT in preparing and running
a very fruitful workshop is gratefully acknowledged. I would also like to thank L.~M.~Sehgal very much for many comments concerning various
topics discussed in this paper. 
\end{theacknowledgments}



\bibliographystyle{aipproc}   



\end{document}